\documentclass[11pt]{article}

\usepackage{amsmath,amssymb,amsfonts,latexsym,amscd,amsthm}
\usepackage{fancyhdr,color,epsfig,setspace,caption,times}
\usepackage[colorlinks=true,plainpages=false,linktocpage,linkcolor=blue,citecolor=green,pagebackref]{hyperref}
\usepackage[letterpaper,left=1.5in,top=1in,bottom=1in,right=1.5in,includeheadfoot,headheight=16.6pt]{geometry}
\usepackage{graphicx}

\usepackage{longtable}
\usepackage{bbm}
\usepackage{dsfont}
\renewcommand{\theequation}{\thesection.\@arabic\c@equation}
\makeatother
\numberwithin{equation}{section}
\begin{document}

\title{Gaming the Quantum}
\author{Faisal Shah Khan, Simon J.D. Phoenix \\ 
Khalifa University, Abu Dhabi, UAE}
\maketitle

\begin{abstract}
In the time since the merger of quantum mechanics and game theory was proposed formally in 1999, the two distinct perspectives apparent in this merger of applying quantum mechanics to game theory, referred to henceforth as the theory of ``quantized games'', and of applying game theory to quantum mechanics, referred to henceforth as ``gaming the quantum'', have become synonymous under the single ill-defined term ``quantum game''. Here, these two perspectives are delineated and a game-theoretically proper description of what makes a multiplayer, non-cooperative game quantum mechanical, is given. Within the context of this description, finding Nash equilibrium in a strictly competitive quantum game is exhibited to be equivalent to finding a solution to a simultaneous distance minimization problem in the state space of quantum objects, thus setting up a framework for a game theory inspired study of ``equilibrium'' behavior of quantum physical systems such as those utilized in quantum information processing and computation. 
\end{abstract}

\section{Introduction}

The modern mathematical approach of Shannon \cite{Shannon} describes information in terms of occurrence of events. The notion of informational content of an event is defined as a quantity that is inversely proportional to the probability with which the event occurs, so that smaller the probability of occurrence of an event the more informational content in the event, and vice versa. The fundamental quantifier of information is therefore a probability distribution, which in the case of a collection of finitely many events is a sequence of real numbers adding up to one, with each number in the sequence representing the probability with which a particular event in the collection occurs.    

Quantum information studies the behavior of information under a quantum mechanical model. As such, the notion of probability distribution is replaced with the notion of quantum superposition followed by measurement. Quantum information and ``classical'' information are fundamentally different, with perhaps the most dramatic difference exhibited by the fact that certain quantum superpositions of two independent quantum objects produce, upon measurement, probability distributions that are impossible to produce by independent classical objects. For example, no flip of two independent classical coins will ever produce the probability distribution $\left(\frac{1}{2},0,0,\frac{1}{2} \right)$ over the four possible states of the two coins; however, certain ``flips'' of two independent quantum coins (or qubits) can give rise to quantum superpositions of the quantum coins called entangled states which, upon measurement, produce the probability distribution above. With its different than usual behavior together with the possibility that such behavior could make contributions to scientific and technological advancement, quantum information continues to be an active area of research. Relatively recent papers such as \cite{Mosca, Ying, Bruss} are good resources for initiating a study of quantum information and its potential benefits to the fields of algorithms, cryptography, computation, and artificial intelligence. 

Quantum games form a relatively new and exciting area of research within quantum information theory. In the theory of quantum games, one typically identifies features of quantum information with those of multiplayer, non-cooperative games and looks for different than usual game-theoretic behavior such as enhanced Nash equilibria. Indeed, it has been established \cite{Eisert,  Benjamin, Landsburg1} that when features of quantum information are introduced in multiplayer non-cooperative games such as Prisoners' Dilemma and other simple two player, two strategy games, Nash equilibrium outcomes can sometimes be observed that are better paying than those available originally. Recent surveys of quantum games can be found in \cite{Guo, Landsburg, Sharif}. In the case of single player games that are modeled by Markovian dynamics, quantum analogues have been constructed that offer insights into the quantization of certain Markov processes and quantum algorithms \cite{Khan1, LeeJohnson1}. Recent developments in representing quantum games using octonions \cite{Aden} and geometric algebras \cite{Chappell, Chappell1} have also provided insights into the behavior of quantum games.   

As noted above, the prevailing research trend in the theory of quantum games involves {\it quantizing} games, that is, introducing features of quantum information theory to games and seeking insightful game-theoretic results such as Nash equilibria that are better paying than those available originally. An opposite approach where one would introduce features of game theory to quantum information does not appear as a prominent research trend in the current quantum game theory literature. Picking up on this deficiency in the literature, we propose here that this latter approach, which we refer to as {\it gaming the quantum}, can potentially produce insightful quantum information theoretic results. To this end, sections \ref{sec:game} and \ref{sec:mixing} together provide a mathematically formal review of non-cooperative games, the fundamental solution concept of Nash equilibrium in such games, and the notion of randomization via probability distributions in these games that gives rise to the so-called  mixed game. Skipping section \ref{sec:gaming the mixture} for the moment, readers will find in section \ref{sec:quantizing a game} a formal treatment of the notion of a quantized non-cooperative game that builds up on the formalism developed in sections \ref{sec:game} and \ref{sec:mixing}, and shows how quantized games are the result of replacing randomization via probability distributions with the higher order randomization via quantum superposition followed by measurement. To delineate the notions of quantizing a game and gaming the quantum, we first refer back to section \ref{sec:gaming the mixture} where the notion of a stochastic game is introduced by suppressing reference to the game that underlies the mixed game of section \ref{sec:mixing}. This delineation is completed in section \ref{sec:gamedquantum} where reference to the game underlying a quantized game is suppressed and game-theoretic ideas are applied directly to the state space of quantum superpositions. This gives rise to the notion of a quantum game that is more general than a quantized game. Further, section \ref{sec:gamedquantum} introduces, for the first time as far as we can tell, a notion of players' preferences over quantum superpositions which is used to construct a novel geometric characterization of Nash equilibrium in quantum games. These ideas are brought together in Theorem 1 and Corollary 1 which connect the study of Nash equilibria in quantum games to a simultaneous minimization problem in the Hilbert space of quantum superpositions. Finally, section \ref{sec:quantum mechanism} proposes an novel synthesis of quantum games and quantum logic circuits by way of an application of Theorem 1 and Corollary 1 to the problem of designing mechanisms for quantum games at Nash equilibrium. 

\section{Non-cooperative Games}\label{sec:game}

Multiplayer, non-cooperative game theory can be described as the science of making optimal choices under given constraints. To this end, introduce a set $O$ of {\it outcomes}, a finite number, say $n$, of individuals called {\it players} with non-identical preferences over these outcomes, and assume that the players interact with each other within the context of these outcomes. Call this interaction between players a {\it game} and define it to be a function $G$ with range equal to the set $O$ and domain equal to the Cartesian product $S_1 \times S_2 \times \dots \times S_n$ with $S_i$ defined to the set of $i^{\rm th}$ players {\it pure strategies}. Call an element of the set $S_1 \times S_2 \times \dots \times S_n$ a {\it play of the game} or a {\it strategy profile}. In symbols, a game $G$ is a function 
$$
G: S_1 \times S_2 \times \dots \times S_n \rightarrow O.
$$ 
Note that a game is distinguished from an ordinary function by the notion of non-identical preferences, one per player, defined over the elements of its range or the outcomes. Also note that it is assumed that players make strategic choices independently, a fact stressed by the fact that a game's domain is the Cartesian product of the pure strategy sets of the players.

Given a game $G$ as above, assume further that the each player has full knowledge of his and his opponents' preferences over the outcomes, of pure strategies available to him and to his opponents, and that each player is aware that he and his opponents are all {\it rational} in that they all engage in a play of the game that is consistent with their respective preferences over the outcomes. In an ideal scenario, players will seek out a play that produces an {\it optimal} (also known as Pareto-optimal) outcome $o$, that is, an outcome that cannot be improved upon without hurting the prospects of at least one player. However, players typically succeed in seeking out a play that satisfies the constraint of their non-identical preferences over the outcomes. Such an outcome has the property that in the corresponding strategy profile (or the play of the game) each player's strategy is a {\it best reply} to all others, that is, any unilateral change of strategy in such a play of the game by any one player will produce an outcome of equal or less preference  than before for {\it that} player. Such a play of a game is called a {\it Nash equilibrium} \cite{Nash}. A Nash equilibrium that happens to correspond to an optimal outcome is called Pareto-optimal or simply an optimal Nash equilibrium.
 
\begin{figure}
\centerline{\includegraphics[bb=1in 3.4in 9in 5in,scale=0.5]{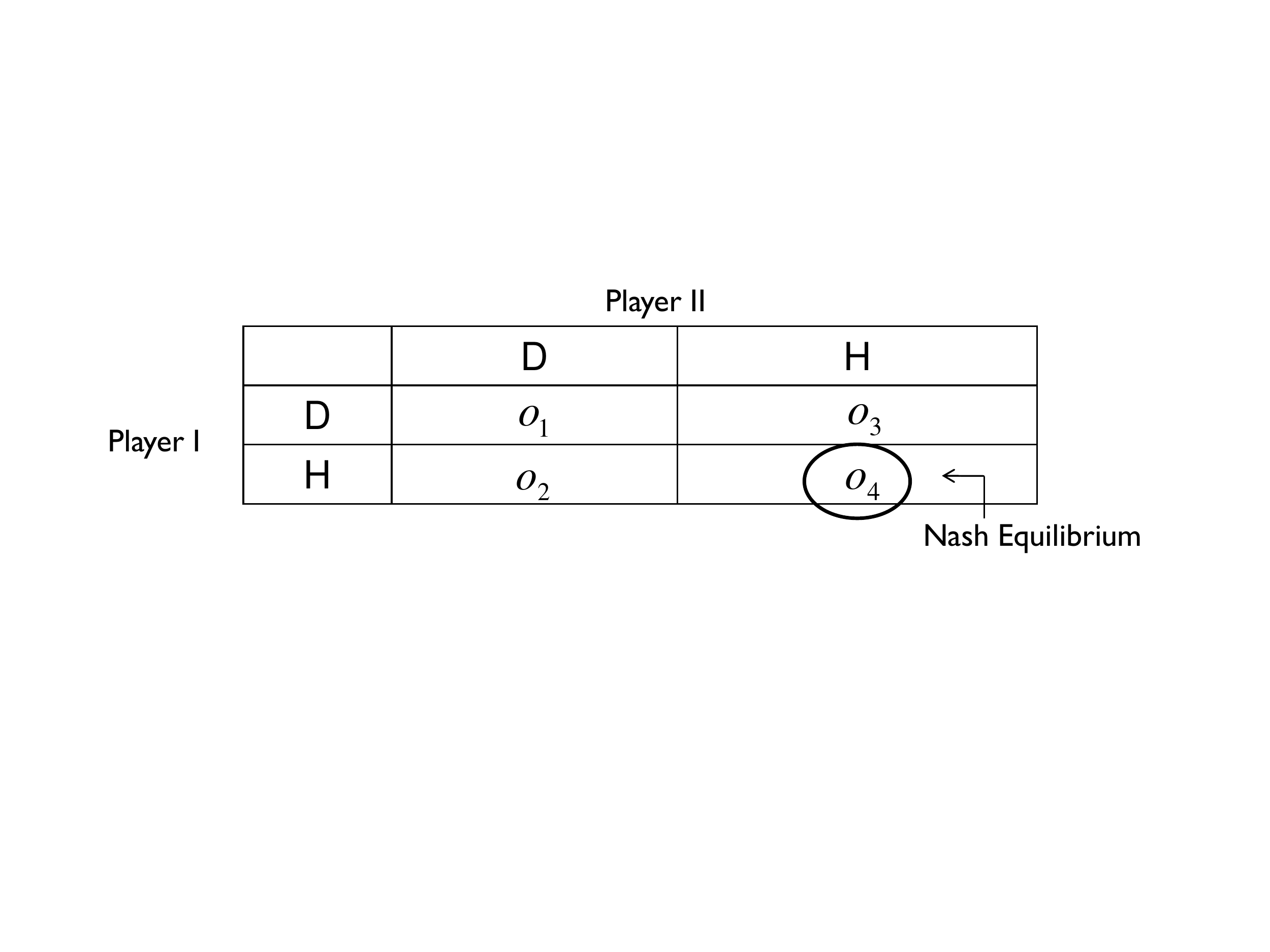}}
\caption{Tabular form of the game $\mathcal{G}$.}
\label{Dove-Hawk}
\end{figure}

Consider as an example the two player, two strategy non-cooperative game $\mathcal{G}$ with the set of outcomes $O=\left\{o_1, o_2, o_3, o_4 \right\}$ and preferences of the players, labeled Player I and Player II, given by
\begin{equation}\label{eqn:I preferences}
{\rm Player \hspace{1mm} I}: o_2 \succ o_1 \succ o_4 \succ o_3
\end{equation}
\begin{equation}\label{eqn:II preferences}
{\rm Player \hspace{1mm}II}: o_3 \succ o_1 \succ o_4 \succ o_2
\end{equation}
with the symbol $\succ$ standing as short-hand for ``more preferred than''. The set of pure strategies of each player is the two element set $\left\{D,H \right\}$ and the game is defined as
$$
\mathcal{G}(D,D)=o_1, \hspace{3mm} \mathcal{G}(D,H)=o_3, \hspace{3mm} \mathcal{G}(H,D)=o_2, \hspace{3mm} \mathcal{G}(H,H)=o_4
$$
Note from the players' preferences that the outcome $o_1$, corresponding to the play $(D, D)$, is optimal. However, the outcome $o_4$ is the unique Nash equilibrium in this game, corresponding to the strategy profile $(H,H)$, because it satisfies the constraint of the players' non-identical preferences over the outcomes in expressions (\ref{eqn:I preferences}) and  (\ref{eqn:II preferences}) in that unilateral deviation from this play of the game by either player results in an outcome that is less preferable. This can be seen more clearly from the tabular format of the game $\mathcal{G}$ in Figure \ref{Dove-Hawk} which shows that the player who unilaterally deviates from the strategy profile $(H,H)$ is left with with an outcome that he prefers less than the outcome $o_4$. 

\subsection{Non-existent or Sub-optimal Nash Equilibria and Game Extensions}\label{sec:unsat}

\begin{figure}
\centerline{\includegraphics[bb=1in 3.4in 9in 5in,scale=0.5]{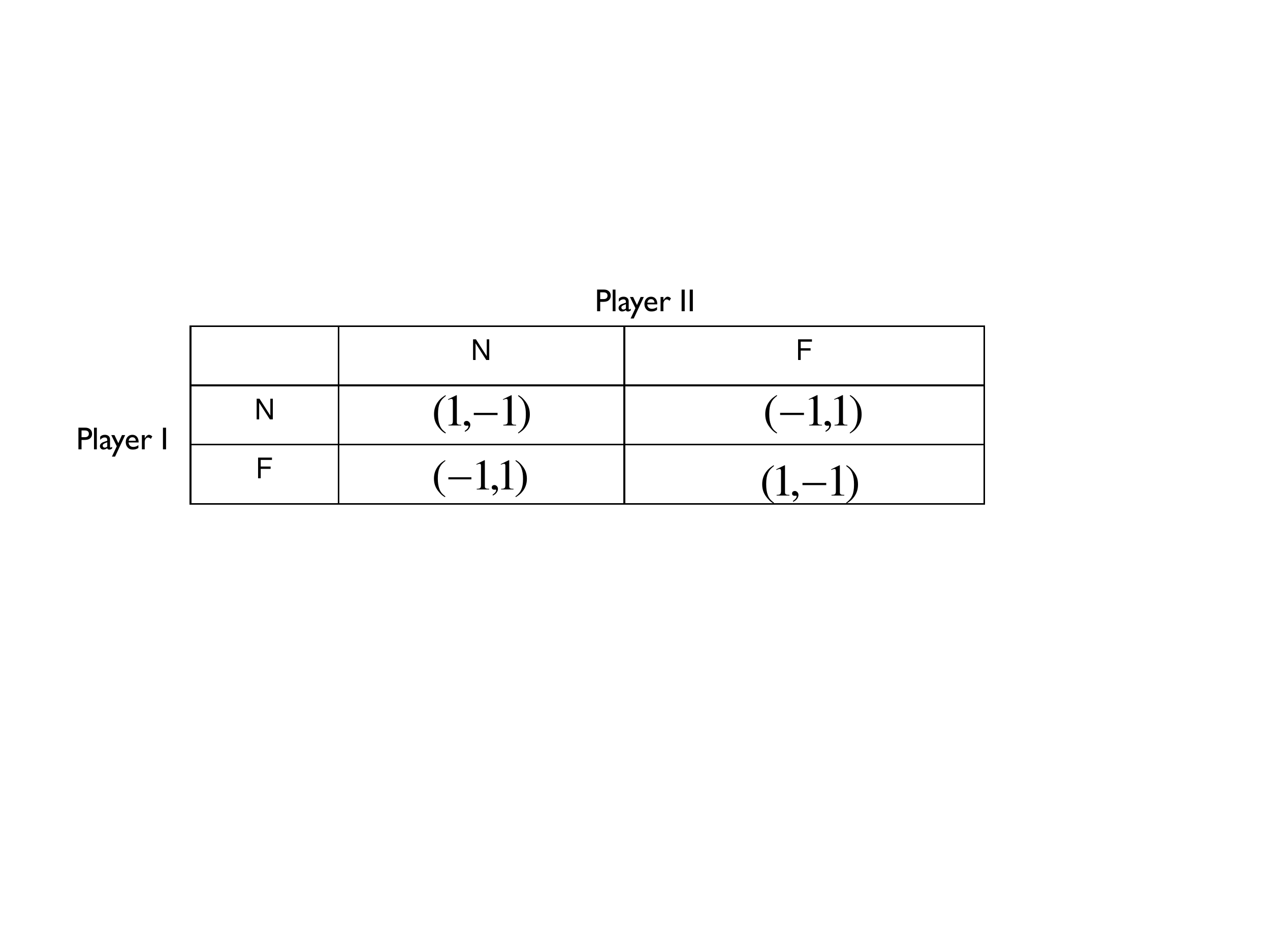}}
\caption{Matching Pennies, a game with no Nash equilibrium.}
\label{MatchingPennies}
\end{figure}

In the game $\mathcal{G}$ above, note that while an optimal outcome exists, it is not the one that manifests as the Nash equilibrium. This situation is considered to be an undesirable solution to the game. Worse game-theoretic situations can occur. For instance, the Matching Pennies game of Figure \ref{MatchingPennies}, where each player prefers $1$ over $-1$ in any play of this game, entertains no optimal outcomes nor any Nash equilibrium! A fundamental problem in game theory is that of finding ways around such undesirable situations. One way this can be achieved is by simple changing the game. For instance, one can talk of changing the players' preferences in the game $\mathcal{G}$ or even changing the function $\mathcal{G}$ itself. However, changing a game can naturally be viewed as a form of cheating. Indeed, a humorous lament betrays this assessment of changing a game as follows: ``When I finally figured out answers to all of life's questions, they changed all the questions!'' On a more practical note, one could argue that changing a game can be costly. For example, the cost of changing a game may be social, such as the one that Ghandi and the inhabitants of the Indian sub-continent incurred in their changing of the British Raj game.  

Is there a way then to overcome undesirable game-theoretic situations where Nash equilibria are non-existent or sub-optimal, without changing the game? An affirmative answer can be found in the form of randomization via probability distributions. Randomization via probability distribution is a time-honored method of avoiding undesirable game-theoretic situations that has been practiced by players since time immemorial in the form of tossing coins, rolling die, or drawing straws, but was formally introduced to game-theory by von Neumann \cite{Neumann}.   

\section{Mixing the game}\label{sec:mixing}

Randomization via probability distributions is introduced in a game formally as an exercise in the extension of the game. Bleiler provides an excellent treatment of this perspective in \cite{Bleiler}. Taking the game $\mathcal{G}$ as an example, we extend its range to include probability distributions over the outcomes by identifying these outcomes with the corners of the simplex of probability distributions over four things, or the 3-simplex $\Delta_3$. Now probability distributions over the outcomes of $\mathcal{G}$ can be formed while at the same time any one of these original outcomes can be recovered by putting full probabilistic weight on that outcome. Formally, the set $O=\left\{o_1, o_2, o_3, o_4 \right\}$ is embedded in the set
$$
\Delta_3=\left\{(p_1, p_2, p_3, p_4):0 \leq p_i \leq 1, \sum_{i=1}^{4}p_i=1 \right\}
$$
via the identification of each outcome $o_i$ with some corner of $\Delta_3$ so that $p_i$ is the probabilistic weight on $o_i$. 

Next, a notion of non-identical preferences of the players over probability distributions is defined. A typical way to define such preferences in game theory is via the notion of {\it expectation}, constructed by assigning a numeric value, one per player, to each of the outcomes such that the assignment respects the preferences of the players over the outcomes. For a given probability distribution, expectation is defined to be 
\begin{equation}\label{eqn:exp}
E(p_1,p_2,p_3,p_4)=p_1a_1+p_2a_2+p_3a_3+p_4a_4 \in \mathbb{R}
\end{equation}
with $a_i$ being the numeric value of the outcome $o_i$. A player will now prefer one probability distribution $P$ over another $Q$ if $E(P) > E(Q)$. The {\it mixed} game $M$ is defined next with domain equal to the Cartesian product of the sets of probability distributions over the pure strategies of the players. In symbols, 
$$
M: \Delta_1 \times \Delta_1 \rightarrow \Delta_3
$$ 
where $\Delta_1$ is the set of probability distributions over the pure strategies of each player and is referred to as the set of {\it mixed strategies} of the players. The mixed game $M$ is defined as
$$
M: \left((p, 1-p), (q, 1-q)\right) \mapsto \left(pq, (1-p)q, p(1-q), (1-p)(1-q) \right)
$$
Figure \ref{Mixedgame} gives a pictorial representation of the construction of a mixed game for the game $\mathcal{G}$. Note that the original game $\mathcal{G}$ can be recovered from the mixed game $M$ by restricting the domain of $M$ to $\left\{D,H \right\} \times \left\{D,H \right\}$. Therefore, even though the mixed game is an entirely different game, the fact that the original game ``sits inside'' it and can be recovered if so desired, game theorists argue that the mixed game is a way around undesirable game-theoretic situations such as sub-optimal Nash equilibria, {\it without changing the game}. 

The existence of Nash equilibria in the mixed game was addressed by John Nash \cite{Nash} who showed that a mixed game with a finite number of players is guaranteed to entertain at least one Nash equilibrium. This powerful result offers a way around the most undesirable situation possible in multi-player non-cooperative game theory, namely, a game without any any Nash equilibria. Moreover, it is sometimes the case that Nash equilibrium in the mixed game are near or fully optimal relative to the original game. 
 
\section{Gaming the Mixture}\label{sec:gaming the mixture}

What can be said about the remains of the mathematical construction employed to produce the mixed game $M$ when the underlying game $\mathcal{G}$ and its domain are removed from consideration? In this case, the function that remains, call it $M'$, maps directly from $\Delta_1 \times \Delta_1$ into $\Delta_3$ and is no longer an extension of $\mathcal{G}$. The corners of each $\Delta_1$ no longer correspond to pure strategies of the players in the game $\mathcal{G}$ and $\mathcal{G}$ is no longer recoverable from $M'$ via appropriate restrictions. However, because the range of the game $\mathcal{G}$, together with players preferences defined over its outcomes, is still intact within $\Delta_3$, a notion of players' preferences over elements of $\Delta_3$ as constructed in section \ref{sec:mixing} still holds. The new function $M'$ can therefore be considered to be a multi-player non-cooperative game in which the set of outcomes is $\Delta_3$ with players' preferences over these outcomes (probability distributions) defined in terms of players' preferences over the corners of $\Delta_3$ via expectation, and the ``pure strategy'' sets of each player equal $\Delta_1$: 
$$
M': \Delta_1 \times \Delta_1 \rightarrow \Delta_3. 
$$   
In other words, $M'$ is the result of ``gaming the mixture'' or an application of ideas from multi-player non-cooperative game theory discussed in section \ref{sec:game} to the stochastic function $M'$. Contrast this with the construction of the mixed game $M$ which can be described as an application of stochastic functions to multi-player non-cooperative game theory. The function $M'$ can appropriately be called a {\it stochastic game}. Note that players' preferences over probability distributions need not be left over artifacts of players' preferences over the outcomes of the now non-existent underlying game $\mathcal{G}$. Players' preferences over probability distributions can always be defined from scratch. 

This idea of casting a given function in a game-theoretic setting by starting with a game extension and then removing the underlying game from consideration is extended to functions used in quantum mechanics in section \ref{sec:gamedquantum}. To this end, a relevant discussion on extensions of games to include quantum mechanics appears in section \ref{sec:quantizing a game} below.

\begin{figure}
\centerline{\includegraphics[bb=1in 1.5in 9in 6.3in,scale=0.55]{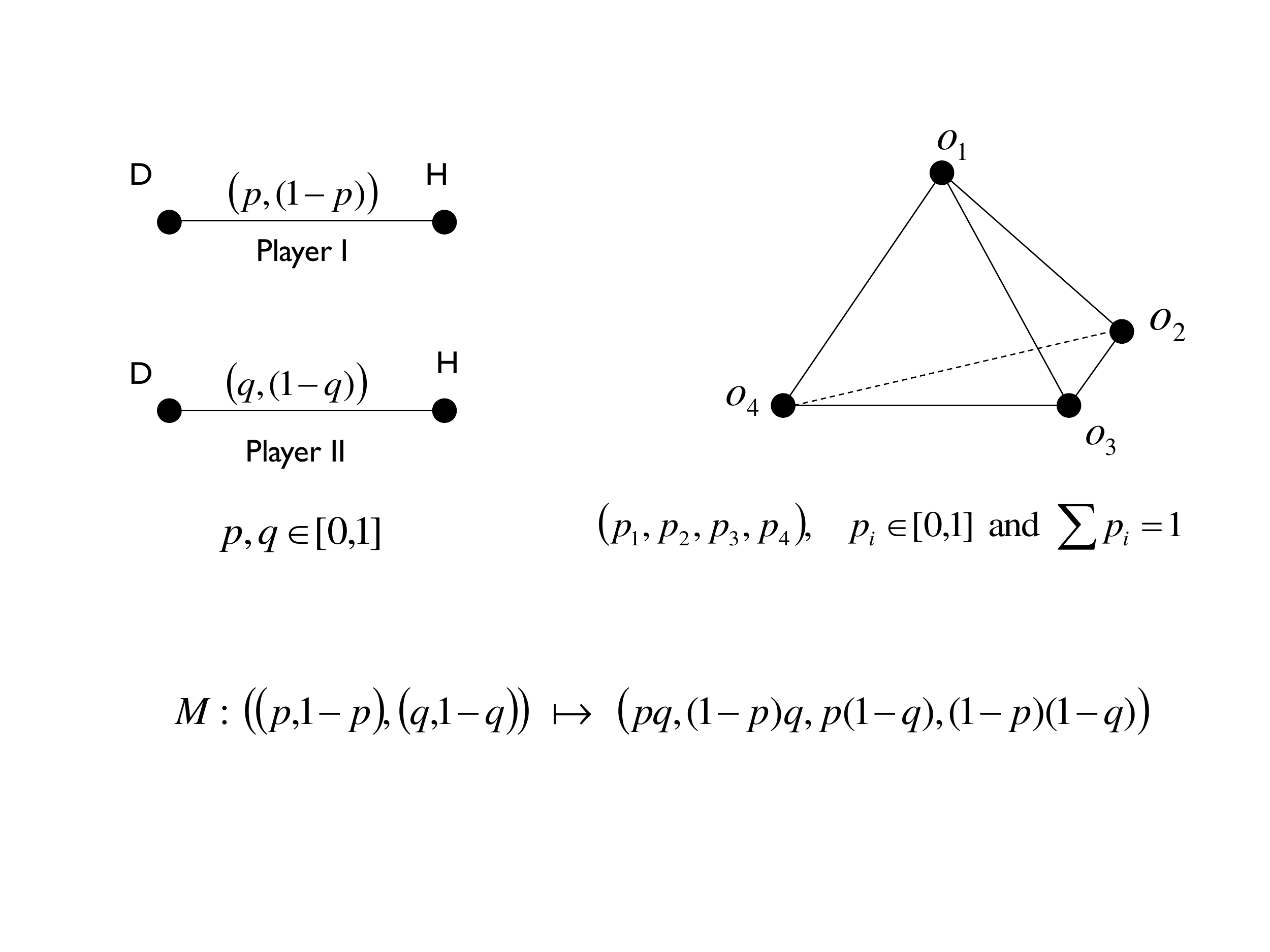}}
\caption{Construction of the mixed game $M$.}
\label{Mixedgame}
\end{figure}

\section{Quantizing the Game}\label{sec:quantizing a game}

Note that because the function $M$ does not map onto $\Delta_3$, the image of the mixed game may not contain probability distributions that are optimal or near-optimal with respect to players' preferences over probability distributions. As such, the mixed game might not entertain Nash equilibria that are better those available originally. This is indeed the case with the game $\mathcal{G}$ of section \ref{sec:game}. In such persistent unsatisfactory game-theoretic situations, game-theorists seek other extensions of the original game.
\begin{figure}
\centerline{\includegraphics[bb=1in 1.5in 9in 6.3in,scale=0.55]{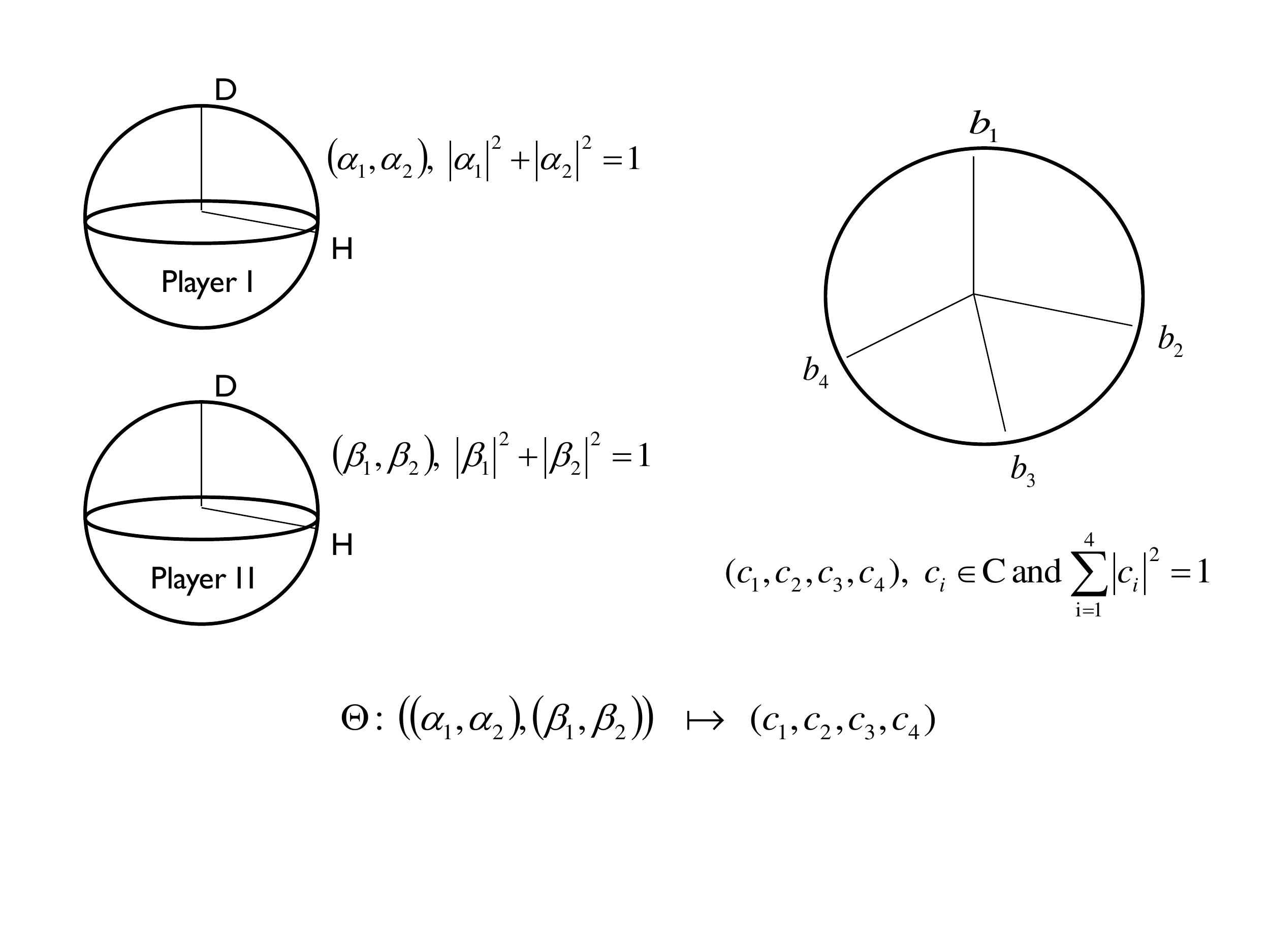}}
\caption{Construction of a quantized game $\Theta$. The quantized game here is in fact the family of quantizations prescribed by Eisert et al. }
\label{Mixedgame}
\end{figure}
Of special relevance here is the extension of a game, formally suggested by Meyer \cite{Meyer}, to included higher order randomization via quantum superpositions followed by measurement. To this end, the outcomes of a game are identified with an orthogonal basis of the space of quantum superpositions.  Mathematically, the space of quantum superpositions is a projective complex Hilbert space. For the game $\mathcal{G}$, a four dimensional projective Hilbert space $\rm{\bf H}_4$ is required with the four outcomes $\left\{ o_1,o_2,o_3,o_4 \right\}$ identified with an orthogonal basis $B=\left\{ b_1, b_2, b_3, b_4 \right\}$ of $\rm{\bf H}_4$. Now quantum superpositions of the outcomes of $\mathcal{G}$ can be formed. Formally, the set $O=\left\{o_1, o_2, o_3, o_4 \right\}$ is embedded in the set 
$$
{\rm{\bf H}}_4=\left\{c_1b_1+c_2b_2+c_3b_3+c_4b_4=(c_1, c_2, c_3, c_4) : \sum_{i=1}^{4}|c_i|^2=1 \right\}
$$
via the identification of of each $o_i$ with an element of the orthogonal basis $B$ so that $c_i$ is the projective complex-valued weight on $o_i$ and $|c_i|^2$ is the square of the norm of the complex number $c_i$. Measurement, denoted here as $meas$, of a quantum superposition $q$ produces a probability distribution in $\Delta_3$ from which the expectation of the outcomes of $\mathcal{G}$ to the players can be computed and the optimality of a quantum superposition can be defined. Note that anyone of the outcomes of the game $\mathcal{G}$ can be recovered by putting full quantum superpositional weight on the basis element corresponding to that outcome and making a measurement. A {\it quantized game} $\Theta$ is defined next, typically as a unitary function from the Cartesian product of sets of quantum superpositions of the players' pure strategies into ${\rm {\bf H}}_4$, with the added property that it reduces to the original game $\mathcal{G}$ under appropriate restrictions. In symbols, 
$$
\Theta: {\rm {\bf H}_2} \times {\rm {\bf H}_2} \rightarrow {\rm {\bf H}_4}
$$
where ${\rm {\bf H}_2}$ denotes the set of quantum superpositions of the strategies of the players, referred to in the literature as quantum strategies of the players. However, because a quantized game is meant to be an extension of the original game $\mathcal{G}$ in a fashion analogous to the mixed game of section \ref{sec:mixing}, we propose that it is more appropriate to refer to ${\rm {\bf H}_2}$ as the set of a player's {\it quantized strategies}. We also point out that the term quantized game in fact refers to an entire family of games, in stark contrast to the term mixed game which refers to a specific function. Depending on the exact nature of $\Theta$, when followed by measurement, the image of this composite map $meas \circ \Theta$ can be larger than the image of mixed game and may contain optimal or near-optimal probability distributions.  

Quantization of games with two players, each having two pure strategies, have been studied extensively with a recent survey of this subject appearing in \cite{Landsburg}. One quantization that underpins most studies of two player, two strategy games is the one proposed by Eisert, Wilkens, and Lewenstein (EWL) \cite{Eisert}. It can be shown that the EWL quantization is the specific family of functions explicitly defined as  
$$
\Theta: \left(\left(\alpha_1,\alpha_2\right), \left(\beta_1,\beta_2\right) \right) \rightarrow \left( c_1, c_2, c_3, c_4 \right)
$$
where 
$$
c_1= (\alpha_1\beta_1+\alpha_2\beta_2)+(\bar{\alpha_1}\bar{\beta_1}+\bar{\alpha_2}\bar{\beta_2})
$$
$$
c_2=-\eta(-\alpha_1\bar{\beta_2}+\alpha_2\bar{\beta_1})+\bar{\eta}(-\bar{\alpha_2}\beta_1+\bar{\alpha_1}\beta_2)
$$
$$
c_3= \bar{\eta}(-\alpha_1\bar{\beta_2}+\alpha_2\bar{\beta_1})-\eta(-\bar{\alpha_2}\beta_1+\bar{\alpha_1}\beta_2)
$$
$$
c_4=-i(\alpha_1\beta_1+\alpha_2\beta_2)+i(\bar{\alpha_2}\bar{\beta_2}+\bar{\alpha_1}\bar{\beta_1}) 
$$
$$
\eta=e^{\frac{i \pi}{4}}=\frac{1}{\sqrt{2}}(1+i),
$$
and focuses on a particular variation of the game $\mathcal{G}$ with specific numeric values replacing the outcomes $o_i$. This is the popular game known as Prisoner's Dilemma. The authors of EWL show that a Nash equilibrium with an expectation equal to the optimal outcome in the original game of Prisoners' Dilemma manifests in the quantized game when the play of the quantized game is restricted to a certain sub-class of quantized strategies. However, when plays of the quantized game consisting of the most general class of quantized strategies are considered, no Nash equilibria manifest! This situation is remarkably different from that of the mixed game where Nash's theorem guarantees the existence of at least one Nash equilibrium. This property of EWL quantization is presented in a more general setting of two player, two strategy games by Landsburg in \cite{Landsburg2} where quaternionic coordinates are utilized to produce the result. 

\section{Gaming the Quantum}\label{sec:gamedquantum}

Just as removing the underlying game $\mathcal{G}$ and its domain from consideration in the construction of the mixed game $M$ in section \ref{sec:mixing} leaves the function $M'$ that can be viewed as an example of applying multi-player non-cooperative game theory to stochastic functions, removing the game $\mathcal{G}$ from consideration in the construction of a quantized game $\Theta$ leaves behind a function $\Theta'$ that can be viewed as an example of an application of multi-player, non-cooperative game theory to quantum mechanics or ``gaming the quantum''. Viewing the function $\Theta'$ as quantum physical function because it maps into the joint state space of quantum mechanical objects, it truly deserves to be called a quantum game. In other words, {\it a quantum game is any function mapping into a projective complex Hilbert space provided a notion of preferences, one per player, is defined over quantum superpositions}. The factors in the domain of a quantum game can now be correctly referred to as the set of {\it quantum strategies} of the players. In a more general setting, the set of quantum strategies can conceivably be any set and need not be restricted to ${\rm {\bf H}}_2$. 

\begin{figure}
\centerline{\includegraphics[bb=1in 1.5in 9in 6.3in,scale=0.55]{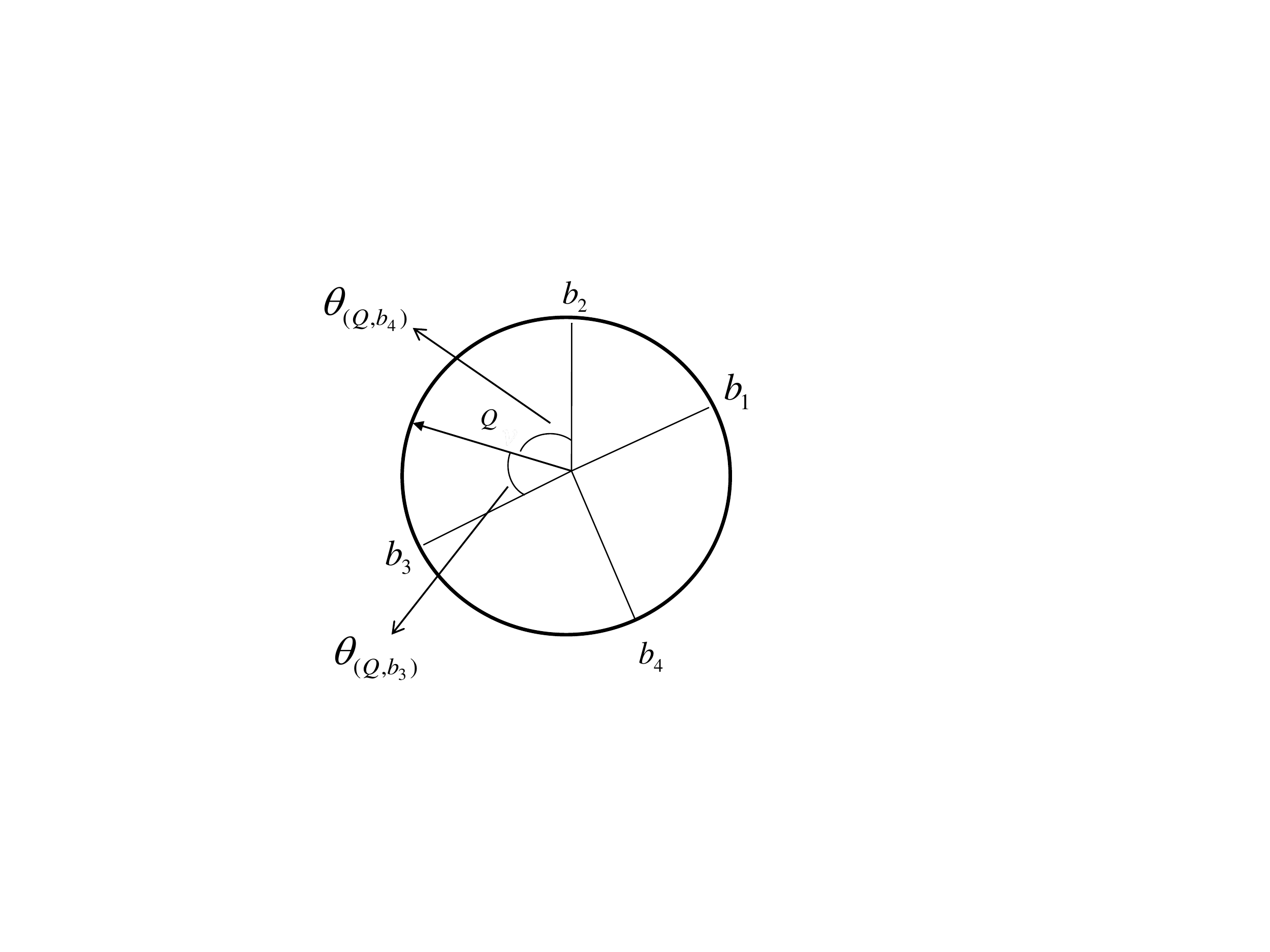}}
\caption{Distance between a quantum superposition $Q$ and elements $b_i$ of an orthogonal basis of ${\rm {\bf H}}_4$.}
\label{gamedquantum}
\end{figure}

While it is certainly valid to utilize post-measurement expectation from quantum superpositions to define players' preferences over quantum superposition, doing so neglects the mathematical structure of the projective complex Hilbert space of quantum superpositions which is richer than that of the simplex of probability distributions. More precisely, the space of quantum superposition entertains a natural notion of distance in terms of its inner-product which can used to define players' preferences via an orthogonal basis of the space of quantum superpositions. Consider ${\rm {\bf H}}_4$ with the orthogonal basis $B=\left\{ b_1, b_2, b_3, b_4 \right\}$ from section \ref{sec:quantizing a game} above as an example and define the players' preferences over the elements of $B$ as 
\begin{equation}\label{eqn:quantumpref}
{\rm Player \hspace{1mm} I}: b_2 \succ b_3 \equiv b_1 \equiv b_4
\end{equation}
\begin{equation}\label{eqn:quantumpref1}
{\rm Player \hspace{1mm}II}: b_3 \succ b_2 \equiv b_1 \equiv b_4
\end{equation}
where the symbol ``$\equiv$'' represents a player's indifference between the two basis elements surrounding the symbol. The choice of preferences in equations (\ref{eqn:quantumpref}) and (\ref{eqn:quantumpref1}) is motivated by the setting of Grover's quantum search algorithm where exactly one element of some database, after identification with an orthogonal basis of some quantum system, is sought out or most preferred and all other elements are less preferable. Preferences like these correspond to the proverbial ``one man's meat is another man's poison'' situation and give rise to strictly competitive games where what is best for one player is the worst for the other(s). From this point of view, quantum search algorithms like Grover's algorithm are strictly competitive quantum games. For strictly competitive games, Nash equilibrium takes on a more restricted nature in the form of a min-max solution where one player attempts to minimizes his maximum possible loss in response to the other player's attempts to maximize his minimum possible gain.   

Returning to the discussion on the mathematical structure of the Hilbert spaces, let $x_1$, $x_2$ $\in {\rm {\bf H}}_4$. The distance between these two elements is given by the angle
$$
{\rm dist}(x_1,x_2)= \theta_{(x_1,x_2)}=\cos ^{-1} \left(|\left\langle x_1, x_2 \right\rangle| \right)
$$
where $\left\langle x_1, x_2 \right\rangle$ is the inner-product of $x_1$ and $x_2$, $|\left\langle x_1, x_2 \right\rangle|$ represents its length or norm, and $\theta \in [0, \frac{\pi}{2}]$. Player I will now prefer one quantum superposition $q$ over another $p$ if $q$ is closer to $b_2$ than $p$ is, that is 
$$
{\rm Player \hspace{1mm} I}: q \succ p \Leftrightarrow {\rm dist}(q, b_2) < {\rm dist}(p, b_2).
$$
Similarly, Player II will prefer one quantum superposition $s$ over another $r$ if $s$ is closer to $b_3$ than $r$ is:
$$
{\rm Player \hspace{1mm} II}: s \succ r \Leftrightarrow {\rm dist}(s, b_3) < {\rm dist}(r, b_3).
$$ 
The notion of Nash equilibrium can now be characterized as a play of a quantum game $\Theta'$ that satisfies the constraints of the players' preferences via this distance notion. Let $Q$ be a quantum superposition corresponding to a play $(x^*,y^*)$ of the quantum game $\Theta'$, that is
\begin{equation}\label{eqn:NE}
\Theta'(x^*,y^*)=Q.
\end{equation}
The quantum superposition $Q$ will be a Nash equilibrium outcome if unilateral deviation on part of any one player from the corresponding play $(x^*,y^*)$ will produce a quantum superposition of lesser preference for {\it that} player than $Q$. Therefore, if Player I deviates from his quantum strategy $x^*$ and instead employs the quantum strategy $x$, then 
\begin{equation}\label{eqn:QI}
\Theta'(x, y^*)=Q_x \hspace{3mm} {\rm with} \hspace{3mm} {\rm dist}(Q_x, b_2) \geq {\rm dist}(Q, b_2).
\end{equation}
Also, if Player II deviates from his quantum strategy $y^*$ and instead employs the quantum strategy $y$, then 
\begin{equation}\label{eqn:QII}
\Theta'(x^*, y)=Q_y \hspace{3mm} {\rm with} \hspace{3mm} {\rm dist}(Q_y, b_3) \geq {\rm dist}(Q, b_3).
\end{equation}
The characterization of Nash equilibrium captured by equations (\ref{eqn:QI}) and (\ref{eqn:QII}) corresponds to a simultaneous distance minimization problem in the Hilbert space ${\rm {\bf H}}_4$, giving the following result: \\

\noindent {\bf Theorem 1}: A  {\em necessary} condition for a play $(x^*, y^*)$ of a two player, strictly competitive quantum game $\Theta'$ to be a Nash equilibrium is that it minimize the distance between its image $Q$ in ${\rm {\bf H}}_4$ under $\Theta'$ and the most preferred basis element of each player in ${\rm {\bf H}}_4$.\\

The theory of Hilbert space \cite{Roman} shows that for a given sub-Hilbert space  $\mathcal{S}$ of a Hilbert space $\mathcal{H}$, there always exists a unique element $s \in \mathcal{S}$ that minimizes the distance between elements of $\mathcal{S}$ and any fixed $h \in \mathcal{H}$. This gives: \\

\noindent {\bf Corollary 1}: A {\it  sufficient} condition for the existence of Nash equilibrium in a strictly competitive quantum game $\Theta'$ is for the image of $\Theta'$ to form a sub-Hilbert space of the ${\rm {\bf H}}_4$. 

\subsection{Designing Quantum games at Nash Equilibrium}\label{sec:quantum mechanism}

Physical implementation of a strictly competitive quantum game at Nash equilibrium is a problem of mechanism design. A mechanism design approach to quantum games has been proposed in \cite{Wu}. Here, we propose a mechanism design approach for studying strictly competitive quantum games at Nash equilibrium based on techniques from quantum logic synthesis. Only the outline of this proposal is discussed below, with a more detailed analysis of this approach deferred to a subsequent publication. 

Continuing with the example developed in the preceding section, start with the orthogonal basis $B$ of ${\rm {\bf H}}_4$ and preferences of Player I and Player II over the elements of $B$ defined as in (\ref{eqn:quantumpref}) and (\ref{eqn:quantumpref1}). Design next a quantum mechanism $\Theta'$ and sets of quantum strategies $X_{\rm I}$, $X_{\rm II}$ for Player I and Player II respectively, so that as per Corollary 1 and Theorem 1, the image of $\Theta'$ is a sub-Hilbert space of ${\rm {\bf H}}_4$ and there exist $x^* \in X_{\rm I}$, $y^* \in X_{\rm II}$ such that $\Theta'(x^*, y^*)=Q$ simultaneously minimizes ${\rm dist}(Q,b_2)$ and ${\rm dist}(Q,b_3)$. 

\begin{figure}
\centerline{\includegraphics[bb=1in 3.4in 9in 5in,scale=0.8]{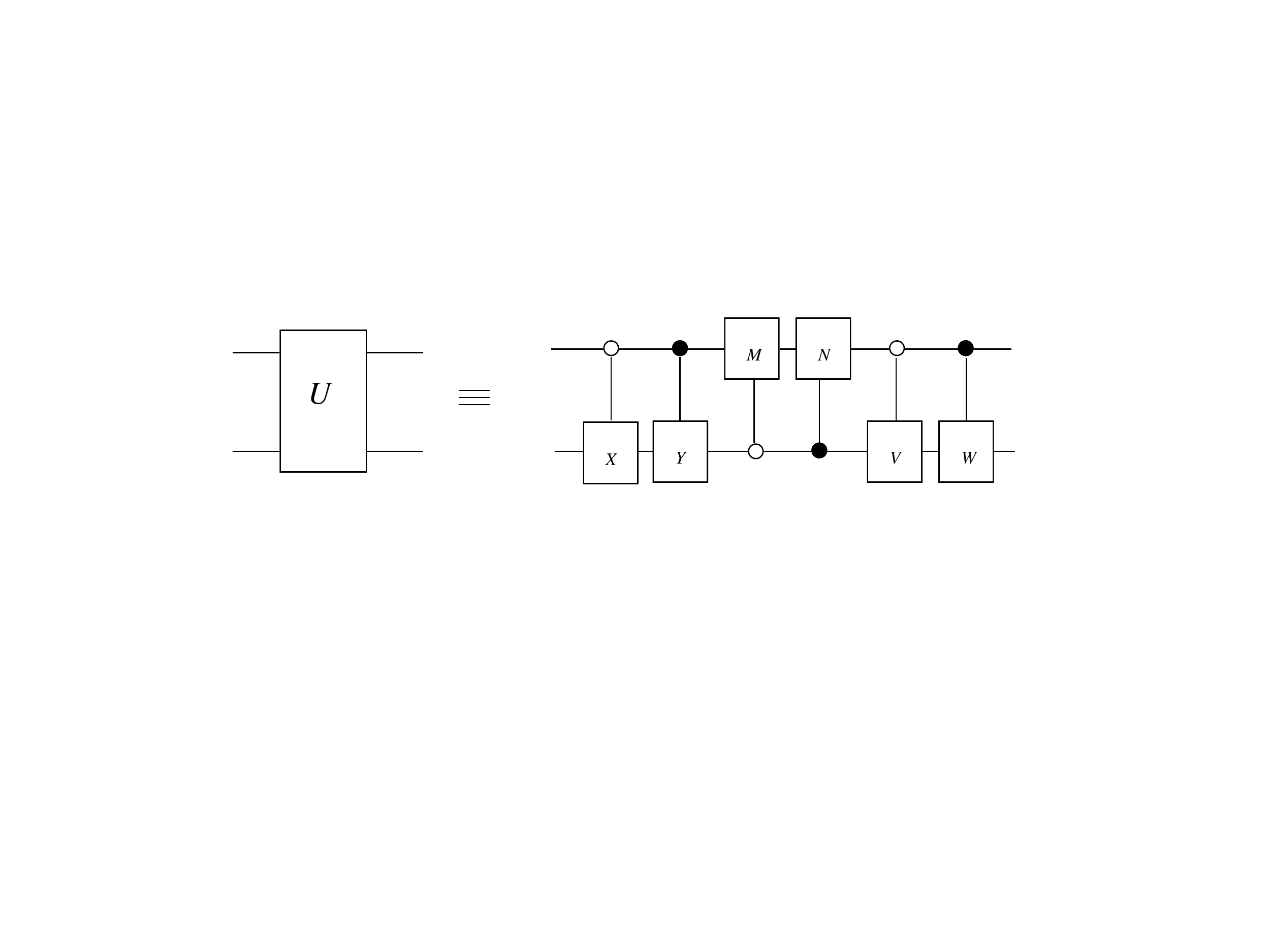}}
\caption{CSD quantum circuit for an arbitrary quantum logic gate $U$. The wires carry qubits and the one qubit gates $X$, $Y$, $M$, $N$, $V$, and $W$ are all controlled either by qubit value $\left|0 \right\rangle$, represented here by the symbol $\circ$, or qubit value $\left|1 \right\rangle$, represented here by the symbol $\bullet$.}
\label{CSDofU}
\end{figure}

Restricting to the case where the quantum game $\Theta'$ is a unitary function mapping into ${\rm {\bf H}}_4$ and $X_{\rm I}=X_{\rm II}={\rm {\bf H}}_2$, the task of designing a quantum game $\Theta'$ at Nash equilibrium equals that of identifying a $4 \times 4$ unitary matrix $U$ and quantum superpositions in each ${\rm {\bf H}}_2$ such that the conditions in both Theorem 1 and Corollary 1 are satisfied. Note the assumption here that the players will make independent quantum strategic choices, sometimes referred to as players' ``local'' actions, although a quantum mechanism where this condition is relaxed is conceivable \cite{Phoenix}. Because the image of any linear function mapping into a finite dimensional Hilbert space is a sub-Hilbert space, and because every unitary function is linear (by definition), the sufficiency condition for a Nash equilibrium of Corollary 1 is satisfied. The problem lies in identifying conditions under which $U$ will satisfy the necessary condition of Theorem 1. A solution to this problem based on quantum logic synthesis is both elegant and appeals to the ultimately physical nature of the problem. 

Viewing $U$ as a quantum logic gate, the mechanism design problem for a quantum game at Nash equilibrium resolves to synthesizing or constructing a circuit for $U$ in terms of {\it universal} quantum logic gates, that is, quantum logic gates which form a circuit that can approximate $U$ up to arbitrary accuracy. It is known that sets of quantum logic gates that map quantum superpositions in ${\rm {\bf H}}_2$ (single qubit) and ${\rm {\bf H}}_4$ (two qubits) are universal \cite{Brylinski}. Moreover, both single and two qubit gates can be implemented practically using most technologies currently available for performing quantum mechanical operations. One technique for quantum logic synthesis, known as the cosine-sine decomposition (CSD), is inspired by the corresponding unitary matrix decomposition technique \cite{Stewart}. The CSD expresses a quantum logic gate as a circuit composed of multiply controlled single qubit gates \cite{Khan, Mottonen, Shende,Tucci}, making the circuit implementable in a practical sense. The CSD quantum logic circuit of an arbitrary $4 \times 4$ unitary matrix $U$, call it $\mathcal{C}$, appears in Figure \ref{CSDofU} where the wires carry qubits and the one qubit gates $X$, $Y$, $M$, $N$, $V$, and $W$ are all controlled either by qubit value $\left|0 \right\rangle$, represented by the symbol $\circ$, or qubit value $\left|1 \right\rangle$, represented by the symbol $\bullet$. The quantum circuit $\mathcal{C}$ can be used to study the existence of specific quantum games or possibly families of quantum games at Nash equilibrium as per Theorem 1. 

\begin{figure}
\centerline{\includegraphics[bb=1in 3.8in 9in 5in,scale=0.8]{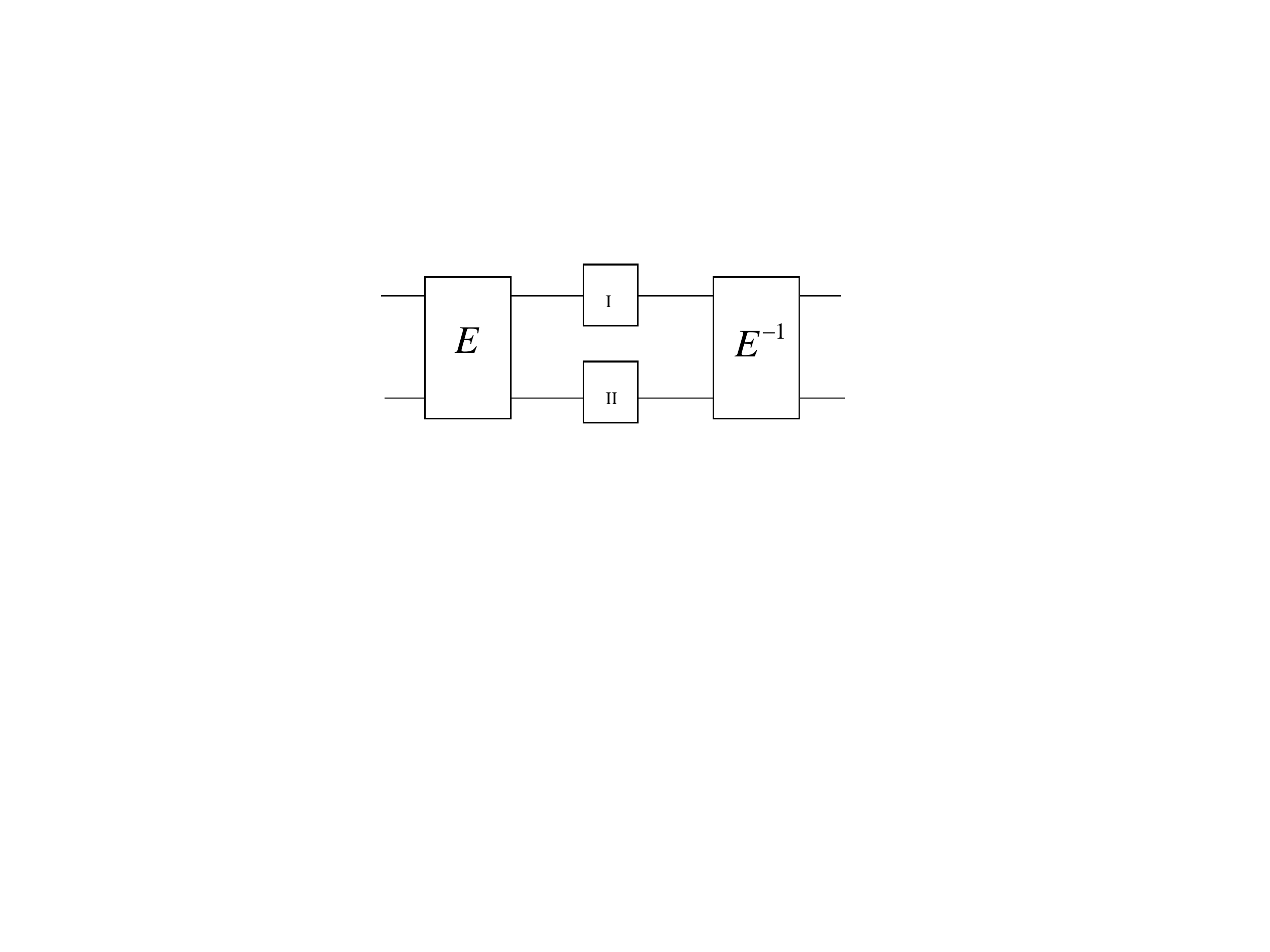}}
\caption{Quantum circuit for EWL quantization of two player, two strategy games like Prisoners' Dilemma. The one qubit quantum logic gates labeled I and II are the players' local actions on their respective qubits. The two qubit gate $E$ entangles the qubits on which the players local actions take place.}
\label{EWLcircuit}
\end{figure}

We point out that $\mathcal{C}$  is a more general construction than the quantum circuit used in the EWL \cite{Eisert} quantization of two player, two strategy games, appearing here in Figure \ref{EWLcircuit}, because it can approximate any quantum logic gate (and circuit) to arbitrary accuracy. As such, it is possible that at a functional level, the EWL quantum circuit can be implemented via some particular instantiation of the quantum circuit $\mathcal{C}$. But to the best of our knowledge, this possibility has not been explored and remains an open question requiring further study. Also note that the EWL quantum circuit is specifically designed so as to recover the underlying classical game such as Prisoners' Dilemma. No such conditions are assumed here for the quantum circuit $\mathcal{C}$. 

\section{Conclusion}

We envision several potential future directions in the area of quantum games. For one, players' preferences given in section \ref{gamedquantum} are strictly competitive in nature and are motivated by a quantum computational and algorithmic context where exactly one outcome, corresponding to a particular calculation or searched item, is the ``correct'' and therefore the most desired outcome of a player. All others are less preferable. On the other hand, another player (or even players) most prefers at least one of the latter. The equilibrium behavior of quantum circuits can potentially be studied from this game-theoretic perspective. Indeed, other preferences that are not strictly competitive in nature may be possible over quantum superpositions, and an entire separate project can be devoted to the study preferences that produce insightful results for quantum games that are not strictly competitive games. 

Further generalization is another possible future direction. One would start with the study of the class of functions into ${\rm {\bf H}}_4$, or indeed into $H_n$ for any $n > N$, that would satisfy the necessary and sufficient conditions for the existence of Nash equilibria. Generalizing further would allow a game-theoretic study of a broader class of functions, culminating with the positive operator valued measurement mapping into infinite dimensional vector spaces with a continuum of basis elements. Note that these generalizations beyond the Hilbert space  ${\rm {\bf H}}_4$ to more interesting mathematical spaces and objects are still grounded in the physics of the quantum, and hence one can still accurately refer to these generalizations as attempts at gaming the quantum. Such studies can potentially produce insightful results in the engineering of control of quantum systems those corresponding to quantum circuits and algorithms. 

\section{Acknowledgment}
The authors gratefully acknowledge useful discussions with Steven Bleiler and Joel Lucero-Bryan.

\end{document}